\documentclass{emulateapj}
\usepackage{graphicx}
\usepackage{color}
\usepackage{amsmath}

\shortauthors{Barclay et al.\ 2012}
\shorttitle{The phase curve of TrES-2b}

\begin{document}

%
\def\ltsima{$\; \buildrel < \over \sim \;$}
\def\lsim{\lower.5ex\hbox{\ltsima}}
\def\gtsima{$\; \buildrel > \over \sim \;$}
\def\gsim{\lower.5ex\hbox{\gtsima}}
\def\K{\emph{Kepler}}
\def\T{TrES-2}
\def\Tp{TrES-2b}

\newcommand{\numax}{\mbox{$\nu_{\rm max}$}}
\newcommand{\Dnu}{\mbox{$\Delta \nu$}}
\newcommand{\dnu}[1]{\mbox{$\delta \nu_{#1}$}}
\newcommand{\muHz}{\mbox{$\mu$Hz}}
\newcommand{\kep}{\mbox{\textit{Kepler}}}
\newcommand{\teff}{\mbox{T_{\rm eff}}}
\newcommand{\msun}{\mbox{$M_{\sun}$}}
\newcommand{\rsun}{\mbox{$R_{\sun}$}}

%

\title{Photometrically derived masses and radii of the planet and star in the TrES-2 system}

\author{
Thomas Barclay\altaffilmark{1,2},
Daniel Huber\altaffilmark{1,11},
Jason F. Rowe\altaffilmark{1,3},
Jonathan J. Fortney\altaffilmark{4},
Caroline V. Morley\altaffilmark{4},
Elisa V. Quintana\altaffilmark{1,3},
Daniel C. Fabrycky\altaffilmark{4,12},
Geert Barentsen\altaffilmark{5},
Steven Bloemen\altaffilmark{6},
Jessie L. Christiansen\altaffilmark{1,3},
Brice-Olivier Demory\altaffilmark{7},
Benjamin J. Fulton\altaffilmark{8},
Jon M. Jenkins\altaffilmark{1,3},
Fergal Mullally\altaffilmark{1,3},
Darin Ragozzine\altaffilmark{9},
Shaun E. Seader\altaffilmark{1,3},
Avi Shporer\altaffilmark{8,10},
Peter Tenenbaum\altaffilmark{1,3},
and Susan E. Thompson\altaffilmark{1,3}
}

\newcommand{\ron}{}
\newcommand{\new}{}

\slugcomment{Accepted for publication in ApJ}

\altaffiltext{1}{NASA Ames Research Center, M/S 244-30, Moffett Field, CA 94035, USA}

\altaffiltext{2}{Bay Area Environmental Research Institute, Inc., 560 Third Street West, Sonoma, CA 95476, USA}

\altaffiltext{3}{SETI Institute, 189 Bernardo Ave, Suite 100, Mountain View, CA 94043, USA}

\altaffiltext{4}{Department of Astronomy and Astrophysics, University of California, Santa Cruz, Santa Cruz, CA 95064, USA}

\altaffiltext{5}{Armagh Observatory, College Hill, Armagh, BT61 9DG, UK}

\altaffiltext{6}{Instituut voor Sterrenkunde, Katholieke Universiteit Leuven, Celestijnenlaan 200 D, B-3001 Leuven, Belgium}

\altaffiltext{7}{Department of Earth, Atmospheric and
Planetary Sciences, Massachusetts Institute of Technology, 77
Massachusetts Ave., Cambridge, MA 02139, USA}

\altaffiltext{8}{Las Cumbres Observatory Global Telescope Network, 6740 Cortona Drive, Suite 102, Santa Barbara, CA 93117, USA}

\altaffiltext{9}{Harvard-Smithsonian Center for Astrophysics, 60 Garden St., Cambridge, MA 02138, USA}

\altaffiltext{10}{Department of Physics, Broida Hall, University of California, Santa Barbara, CA 93106, USA}
\altaffiltext{11}{NASA Postdoctoral Program Fellow}
\altaffiltext{12}{Hubble Fellow}


\begin{abstract}
We measure the mass and radius of the star and planet in the \T{} system using 2.7 years of observations by the \K{} spacecraft. The light curve shows evidence for ellipsoidal variations and Doppler beaming on a period consistent with the orbital period of the planet with amplitudes of $2.79^{+0.44}_{-0.62}$ and $3.44^{+0.32}_{-0.37}$ parts per million (ppm) respectively, and a difference between the day and night side planetary flux of $3.41^{+0.55}_{-0.82}$ ppm. We present an asteroseismic analysis of solar-like oscillations on \T{}A which we use to calculate the stellar mass of $0.94\pm 0.05 M_{\odot}$ and radius of $0.95\pm 0.02 R_{\odot}$. Using these stellar parameters, a transit model fit and the phase curve variations, we determine the planetary radius of $1.162^{+0.020}_{-0.024} R_{\textrm{Jup}}$ and derive a mass for \Tp{} from the photometry of $1.44\pm 0.21 M_{\textrm{Jup}}$. The ratio of the ellipsoidal variation to the Doppler beaming amplitudes agrees to better than 2-$\sigma$ with theoretical predications, while our measured planet mass and radius agree within 2-$\sigma$ of previously published values based on spectroscopic radial velocity measurements. We measure a geometric albedo of $0.0136^{+0.0022}_{-0.0033}$ and an occultation (secondary eclipse) depth of $6.5^{+1.7}_{-1.8}$ ppm which we combined with the day/night planetary flux ratio to model the atmosphere of \Tp{}. We find an atmosphere model that contains a temperature inversion is strongly preferred. We hypothesize that the \K{} bandpass probes a significantly greater atmospheric depth on the night side relative to the day side.

\end{abstract}


\section{Introduction}
The two most productive methods for discovering exoplanets have been surveys hunting for periodic changes in radial velocity motion seen in stellar spectra \citep[e.g.][]{vogt00,mayor09} and photometric searches for planetary transits \citep[e.g.][]{bakos02,alonso04,mccullough05,pollacco06}. From the radial velocity method the planet's minimum mass ($M\sin{i}$) can be derived while the transit method allows for the measurement of the planetary radius and the orbital inclination. Neither technique can give the planet's density when used alone and knowledge of the stellar radius and mass are required for the radial velocity and transit methods, respectively.

Several dozen transiting planets have a measured density from follow-up radial velocity observations (both \url{http://exoplanet.eu} [\citealt{schneider11}] and \url{http://exoplanets.org} [\citealt{wright11}] provide up to date lists of planetary characteristics), but this requires a significant amount of telescope time and the use of high precision spectrographs. Until recently, no planet has had a mass derived from photometry alone. The first planets with masses measured from photometry were found through observations of transit timing variations \citep[e.g.][]{lissauer11, ford12a, steffen12a, fabrycky12, ford12b}. However this technique is limited to systems with multiple dynamically interacting planets.

The analysis of the light curves of transiting exoplanet host stars using space based photometers have enabled the measurement of ellipsoidal variations \citep[a recent review is given in][]{mazeh08} and Doppler beaming caused by transiting planets from which the planetary mass can be determined \citep{hills74,maxted00,loeb03,zucker07}. Ellipsoidal variations are periodic flux variations resulting from a orbiting body raising a tide on the host star. The star is therefore non-spherical and varies in brightness as a function of the visible surface area. The dominant period of this signal will be half the orbital period of the planet. Doppler beaming is a combination of a bolometric and a bandpass dependent effect. As a star moves due to the gravitational pull of a companion, the angular distribution of stellar flux will be beamed in the direction of the star's velocity vector. This effect is bolometric and results in an observed periodic brightness change proportional to the star's radial velocity. The bandpass dependent effect is a periodic red/blue shift in the spectral energy distribution of the star which results in the measured brightness of the star changing as the flux falling within the bandpass changes. The magnitude of this effect depends not only on the radial velocity of the star but also on the stellar spectrum and the observational bandpass.

The number of stars for which either planetary induced ellipsoidal variation or Doppler beaming have been observered can be counted on one hand. \citet{mazeh10} observed ellipsoidal variations and Doppler beaming in the phase curve (the complete flux time-series of the star folded on the planetary orbital period) of CoRoT-3, the host to a 22 $M_{\textrm{Jup}}$ planet/brown dwarf. More recently, \K{} observations of HAT-P-7 \citep{welsh10,mislis12a} and KOI-13.01 \citep{shporer11,mislis12b,mazeh12} have been used to derive planet masses photometrically. Using the whole phase curve has also been shown to be a promising method for confirming the planetary nature of sources without the need for follow-up observations \citep{quintana12}.

Light from the planet itself also contributes to the out-of-transit phase curve profile of host stars. The optical light from a cool planet is mainly owing to scattered starlight from the host star but hotter planets, such as the population of hot Jupiters, also contribute significant thermal emission at both infra-red and optical wavelengths. From the ratio of a planet's flux (the combination of thermal emission and scattered light) to the stellar flux measured at full phase, the geometric albedo can be computed. The optical planet-to-star flux ratio has been measured for a small number of systems \citep[e.g.][]{snellen09,welsh10,demory11,fortney11,desert11c,desert11d} and robust upper limits have been placed on others \citep{rowe08,kipping11}. Combining a measurement of an occultation (secondary eclipse) depth and the star-to-planet flux ratio can help determine the planet's night and day side equilibrium temperatures \citep{harrington06}, atmospheric composition \citep{rowe06}, heat redistribution \citep{showman09} and even changing exoplanet weather patterns.

The study of exoplanets is often limited by the difficulty in determining stellar masses and radii from spectroscopy alone due to the strong degeneracies between effective temperature, surface gravity and metallicity \citep[e.g.][]{torres12}. Uncertainties on a spectroscopically derived radius can be at the level of 50\% \citep{basu12}. Asteroseismic observations of solar-like oscillations can be used to reduce these uncertainties to less than 5\% \citep[e.g.][]{stello09,chaplin11,carter12}.

In this work we analyze the photometric time series observations of the planet host star \T{}A\footnote{In an attempt to avoid confusion, we use the designation TrES-2 to refer to the star system, TrES-2A to refer to the star and TReS-2b to refer to the planet.} (listed in the Kepler Input Catalog with identification number 11446443) made using the \K{} spacecraft. We simultaneously fit transit, occulation, and phase curve effects to determine the radius of \Tp{}\footnote{This source has the Kepler Object of Interest designation KOI\nobreakdash-1.01} and derive the mass of the planet by measuring the amplitude of Doppler beaming and ellipsoidal variations.  \T{} is an ideal target for phase curve analysis because (a) the star is relatively bright ($V = 11.4$), (b) it has been observed by \K{} in short cadence mode since the beginning of the mission\footnote{In short cadence mode the sampling is approximately 1 min as opposed to the standard \K{} sampling of $\sim$30 min \citep{gilliland10}.}, (c) we detect solar-like oscillations on the star, (d) we see very little stellar activity, (e) the orbit of \Tp{} is very close to circular \citep{husnoo12}, (f) no variations in the transit times of \Tp{} are detected \citep{steffen12}, and (g) a detection of light from the planet has previously been claimed (indeed, this planet has previously been reported to be the darkest known exoplanet) \citep{kipping11}.


\section{Observations of TrES-2}
\begin{figure*}
\begin{center}
\includegraphics[width=\textwidth]{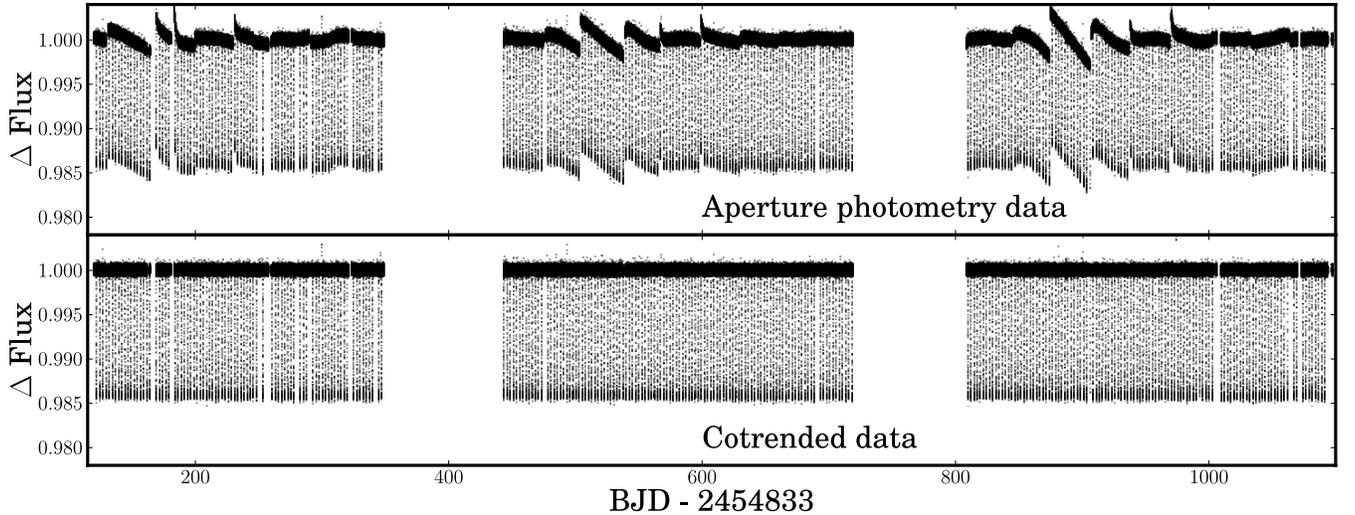}
\caption{The Simple Aperture Photometry data (called SAP\_FLUX in the \K{} FITS files) is shown in the upper panel for available quarters up to Quarter 11. The data from which a combination of linear trends has been removed using the cotrending basis vectors is shown in the lower panel. Data has been normalized to the median for each quarter individually. The gaps in data are owing to the star falling onto Module 3 of the spacecraft photometer during Quarters 4 and 8. Module 3 failed during Quarter 4. The regular pattern of transits can clearly be seen. The short sections of data where transits appear to be `missed' are because no data we collected during these times owing to events such as telescope Earth-points and safe modes. } 
\label{fig:fulllc}
\end{center}
\end{figure*}

The \K{} spacecraft was launched in 2009 with the goal of discovering transiting exoplanets \citep{borucki10,koch10}.
The planet \Tp{} is one of only three to be discovered before the launch of \K{} \citep{odonovan06}. These three previously known exoplanets have been observed in short cadence mode \citep{gilliland10} since the start of the mission. We use the Quarters 0-11 short cadence observations (where the short cadence sampling rate is 58.85 s) of \T{} which span a total of 978 days. No data is used for Q4 and Q8 owing to star falling on a failed CCD during these quarters.
We used simple aperture photometry data (shown in the top panel in Figure~\ref{fig:fulllc}) as opposed to time series which have undergone pre-search data conditioning (PDC) as currently available short-cadence PDC data can distort astrophysical signals on timescales comparable to the orbital period of \Tp{} \citep{stumpe12,smith12}. We removed instrumental signals by fitting cotrending basis vectors to the time series of each quarter individually using the \textsc{PyKE} Kepler community software\footnote{\textsc{PyKE} is available from \url{http://keplergo.arc.nasa.gov/PyKE.shtml}.}. Details of the cotrending basis vectors are given in the Kepler Data Processing Handbook \citep{christiansen12} and their use described in \citet{barclay12} and \citet{kinemuchi12}.
The cotrending basis vectors are only calculated for \K{} long cadence data (29.4 min sampling) so we interpolated onto the short cadence time-stamps using cubic splines. The light curve after the application of the cotrending basis vectors is shown in the lower panel of Figure~\ref{fig:fulllc}.

In order to remove quarter-to-quarter discontinuities we normalized each quarter to the median. Doing this is appropriate because contamination from background sources is very low for \T{} (typically less than 2\%\footnote{Contamination values for each source can be found on the data search website hosted by MAST.}). When analyzing transit photometry it is common for data to be high-pass filtered in order to remove intrinsic stellar variability. However, we found this unnecessary because \T{}A is intrinsically fairly quiet on timescale of days -- there is little discernible variability owing to star-spots. Filtering has the potential to damp the signals we are interested in and so being able to avoid filtering is highly preferable.

\section{The stellar parameters of TrES-2A} \label{sec:star-para}
The stellar properties of \T{}A have so far mainly been studied using high-resolution spectroscopy. 
\citet{sozzetti07} combined constraints on the stellar density from the transit fit by \citet{holman07} 
to derive surface gravity, $\log g=4.44\pm0.02$, effective temperature, $T_{\rm eff}=5850\pm50$\,K 
and metallicity, $\rm{[Fe/H]}=-0.15\pm0.1$, while \citet{ammler09} derived $\log g=4.3\pm0.1$, 
$T_{\rm eff}=5795\pm73$\,K and $\rm{[Fe/H]}=0.06\pm0.08$ using spectroscopy only. 
Additionally, \citet{southworth09} used 
$T_{\rm eff}$ and $\rm{[Fe/H]}$ from \citet{sozzetti07} together with empirical mass-radius relationship
constraints from radial velocity measurements and the transit light curve 
to determine a mass and radius of 
$M/M_{\sun}=0.96\pm0.08$ and $R/R_{\sun}=0.98\pm0.05$, respectively.

\begin{figure}
\begin{center}
\resizebox{\hsize}{!}{\includegraphics{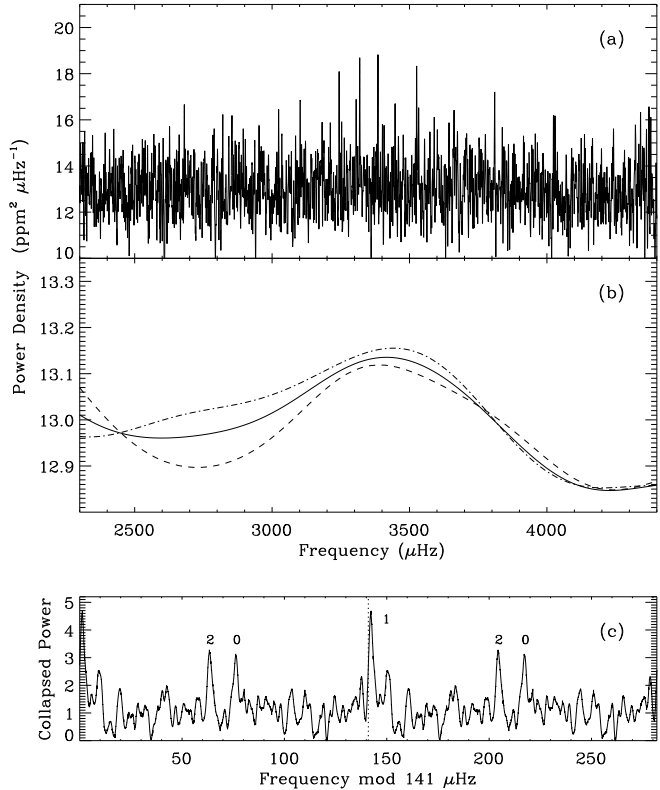}}
\caption{\textit{(a)} The power spectrum of the Kepler Q0--Q11 time series of TrES-2A 
smoothed with a Gaussian filter of full-with at half maximum of $0.5\,\muHz$ and centered 
on the power excess due to solar-like oscillations. 
\textit{(b)} Power spectrum shown in panel (a) but heavily smoothed with a 
full-with at half maximum of 4\Dnu\ (solid line), compared to an equally smoothed 
power spectrum calculated using the first half (dashed line) and second half 
(dashed-dotted line) of the dataset.
\textit{(c)} The power spectrum shown in panel (a) 
folded on large frequency separation between $2900-3800\,\muHz$ and binned. Plotted is 
the sum of the power in each bin. Numbers in the plot 
denote the identification of the spherical degree of each of the three visible ridges. 
The folded plot is shown twice for clarity. The dotted line marks the point where the 
plot repeats itself.}
\label{fig:ps}
\end{center}
\end{figure}

TrES-2A was among the first exoplanet host stars in the Kepler field 
to be analyzed using asteroseismology. The initial analysis by \citet{cd10}, 
based on the first 40 days of data, yielded a tentative detection of equally spaced peaks 
which was found to be in rough agreement with the properties derived by \citet{sozzetti07}.
To improve the stellar properties, we 
have analyzed all available short-cadence data up to Q11 using the method described by \citet{huber09}. 
In summary, the method employs a frequency-resolved autocorrelation to 
locate the excess power due to solar-like oscillations. After correcting the power 
spectrum for contributions due to granuation, the large frequency separation 
(the average spacing between oscillation modes of the same spherical degree and consecutive 
radial order) 
is calculated by fitting a Gaussian function to the highest peak of the power spectrum 
autocorrelation centered on the frequency of maximum power. We note that the method by 
\citet{huber09} has been thoroughly tested against independent methods 
\citep{hekker11,verner11} and has been extensively applied for the analysis of single stars 
and large ensembles of stars observed with Kepler 
\citep[see, e.g,][]{huber11b,howell12,silva12}.

Our analysis yielded a clear detection of  excess power near 3400\,\muHz, consistent with solar-like oscillations. 
Figure \ref{fig:ps}(a) displays the power spectrum centered around the power 
excess. Note that we have checked the frequency range for known artifacts in Kepler 
short-cadence data \citep[see, e.g.,][]{gilliland10b} and found no corresponding peaks in 
our data.
The characteristic regular spacing indicative of p-mode oscillations is visible. 
We find $\Dnu=141.0\pm1.4\,\muHz$, with uncertainties estimated using Monte-Carlo simulations. 
Figure \ref{fig:ps}(c) shows the power 
spectrum shown in Figure \ref{fig:ps}(a) after folding it on a frequency of 
$141\,\muHz$ and summing up all power between $2900-3800\,\muHz$, which roughly 
corresponds to the frequency range where oscillation modes are visible.
There are three peaks in the folded power spectrum which we identify as oscillation modes of 
spherical degree $\ell=0, 1$ and 2. 
The measured small frequency separation, the amount by which $\ell=2$ modes are offset from 
$\ell=0$ modes, from the folded spectrum is $\dnu{02}=12.7\,\muHz$, 
which is fully consistent with a Sun-like star \citep[e.g.,][]{cd88,white11}. 
The S/N is not high enough to a precisely constrain the frequency of maximum power \numax\, and 
it has hence not been used in the remainder of this analysis.

To test the significance of the detection, we first divided the 
data in two parts and calculated the power spectrum of each dataset.
The result in shown in Figure \ref{fig:ps}(b), which shows the 
power spectra after heavily smoothing them with a Gaussian filter with a FWHM = $4\Dnu$ 
to suppress the stochastic nature of the signal \citep[see][]{kjeldsen08}.
The excess power at $\sim$\,3400\muHz\ is visible in both datasets, confirming that the 
excess power in the combined dataset is present in independent parts of the data. 
We furthermore performed simulations by 
generating time series with white noise corresponding to the time-domain scatter of 
the original data, and by randomly shuffling the original data with replacement. For 
each simulated dataset we performed the same analysis as for the original 
data. The results showed that the $l=1$ peak in the collapsed 
power spectrum (which primarily constrains the large separation) is detected at a level of 
$4.2\sigma$, corresponding to a $>99.99\%$ probability that the peak is not due to 
random noise. These tests, combined with the fact that the  
large frequency separation, small frequency separation, and location of the power excess 
are fully consistent for a solar-type star, 
give us confidence that the detected asteroseismic signal is robust.

The large separation can been shown to be 
directly related to the mean density of the star \citep{ulrich86}:

\begin{equation}
\frac{\Delta\nu}{\Delta\nu_{\sun}} \approx \sqrt{\frac{\rho_{*}}{\rho_{\sun}}} \: .
\label{equ:dnu}
\end{equation}

\noindent
Equation (\ref{equ:dnu}) has been tested both theoretically 
\citep[e.g.,][]{stello09b,white11} and empirically \citep[e.g.,][]{miglio11,huber12}.
Using our measured value above we arrive at a mean stellar density of $\rho_{\star}=1.53\pm0.03$ g\,cm$^{-3}$ 
for \T. To estimate a full set of 
stellar parameters, we adopt $T_{\rm eff}$ and [Fe/H] from 
\citet{sozzetti07} and compare our constraints with a dense grid of quadratically interpolated BaSTI 
stellar evolutionary tracks \citep{basti}. 
We start by identifying the maximum likelihood model assuming Gaussian likelihood functions for $\rho_{\star}$, $T_{\rm eff}$ and [Fe/H] \citep[e.g.][]{basu10, kallinger10}. This procedure is then repeated $10^5$ times using values for $\rho_{\star}$, $T_{\rm eff}$ and [Fe/H] drawn from random distributions with standard deviations corresponding to the measurement uncertainty in each parameter. The final modeled parameters and uncertainties are calculated as the median and 84.1 and 15.9 percentile of the resulting distributions. These stellar parameters are listed in Table 1. 

\begin{figure}
\begin{center}
\resizebox{\hsize}{!}{\includegraphics{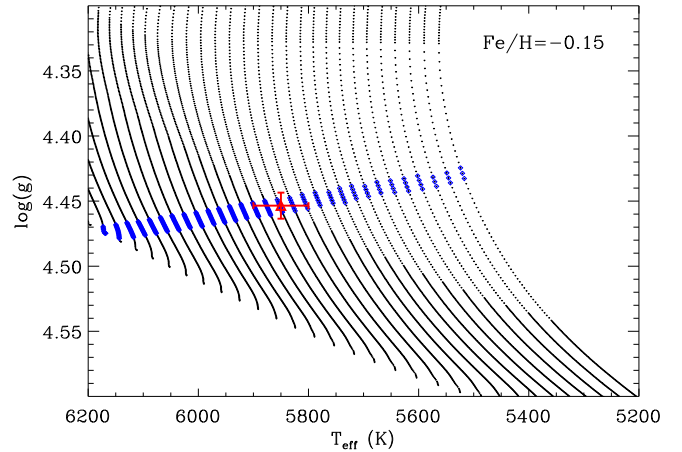}}
\caption{Surface gravity versus effective temperature for 
quadratically interpolated BaSTI evolutionary tracks with masses 
from $0.8-1.1$\msun\ in steps of 0.01\msun\ 
for the metallicity given by \citet{sozzetti07}. 
Blue lines mark all models that satisfy the 1-$\sigma$ constraint on the stellar 
density from asteroseismology. The red triangle shows derived best-fitting surface gravity with 
the effective temperature found by \citet{sozzetti07}.}
\label{fig:hrd}
\end{center}
\end{figure}

Figure \ref{fig:hrd} shows evolutionary tracks in a $\log g-T_{\rm eff}$ plane 
for the metallicity found by \citet{sozzetti07}, illustrating the density 
constraint from the solar-like oscillations in the blue and showing our best-fitting model as a red triangle with uncertainties. 
We note that the uncertainties on the stellar properties stated in Table 1 do not include contributions from uncertainties due to different model grids, which have been shown to be on the order of 2\% in radius and 5\% in mass \citep[e.g.][]{howell12}.

\begin{table}
\begin{center}
\caption{Stellar properties of TrES-2}
\vspace{0.1cm}
\begin{tabular}{l c c}        
\hline         
Property  	& 	Value		& Reference \\  
\hline
$T_{\rm eff}$ (K) 		&	$5850\pm50$		& \citet{sozzetti07}			\\
$\rm [Fe/H]$ (dex)			&	$-0.15\pm0.1$	& \citet{sozzetti07}			\\
$\rho$ (g cm$^{-3}$)	&	$1.53\pm0.03$	& this work			\\
$M$ (\msun)				&	$0.94\pm0.05$	& this work		\\
$R$ (\rsun)				&	$0.95\pm0.02$	& this work 	\\
$\log g$ (dex)			&	$4.45\pm0.01$	& this work		\\
Age (Gyr)				&	$5.8\pm2.2$		& this work	\\
\hline
\end{tabular} 
\label{tab:stellar} 
\end{center}
\end{table}

The stellar properties from our analysis are in reasonably good agreement with the values from 
\citet{sozzetti07} and \citet{southworth09}.
Our analysis presented here gives a direct measurement of the stellar density, 
resulting in refined stellar parameters which are independent of the transit modeling.
We note that a more detailed asteroseismic analysis using individual frequencies,
including constraints from the small frequency separations, can be expected to further 
improve the stellar parameters (in particular the age) of TrES-2A.

\section{Flux time series fitting of TrES-2b} \label{sec:transit-para}
We fit a \citet{mandel02} transit model to the light curve of
\T{} using a 4-parameter non-linear limb darkening law with
coefficients interpolated from \citet{claret11} using ATLAS models, where the interpolation is trilinear in $T_{\rm eff}$ and [Fe/H] and $\log g$. The four limb darkening coefficients are given in Table~\ref{tab:transit}. We parameterize the \citet{mandel02} model such that we fit for the orbital period (P), time of transit ($T_{0}$), impact
parameter (b), mean stellar density ($\rho_{\star}$), scaled planet radius ($R_{p}/R_{\star}$), $e \cos{\omega}$, $e \sin{\omega}$ (where e is eccentricity and $\omega$ is the periastron angle) and
occultation (secondary eclipse) depth. 

In addition, we simultaneously fit for the amplitude of the observable effects caused by ellipsoidal variations, Doppler beaming and reflection/thermal emission from the planet using a modification to the method outlined in \citet{faigler11} as a function of orbital phase, $\phi$. The orbital phase is defined as
\begin{equation}
\phi = \frac{(t-T_{0}) + P}{P} - \lfloor\frac{(t-T_{0}) + P}{P}\rfloor,
\end{equation}
where $P$ is the orbital period of the \Tp, $t$ is time and $T_{0}$ is the time of the first transit. The expression $\lfloor x \rfloor \equiv \mathrm{floor}(x)$ where the `floor' function rounds down to the nearest integer. The phase $\phi$ runs from 0 to 1 and mid-transit occurs at $\phi=0$.  
The change in brightness of the system as a function of phase can be described as
\begin{equation}
\begin{split}
\frac{\Delta F}{F} = f - A_{e} \cos{(4 \pi \phi - l)} + A_{b}\sin{2\pi\phi} &\\
- A_{r}\frac{\sin{z} + (\pi - z)\cos(z)}{\pi}
\end{split}
\end{equation}
where the amplitude coefficients $A_{e}$, $A_{b}$ and $A_{r}$ are owing to the effects of ellipsoidal variations, Doppler beaming and reflection/thermal emission. $f$ is an arbitrary photometric zero-point in flux which is needed because we do not precisely know the flux zero-point before we fit a model. $l$ is a phase lag in the ellipsoidal variations where a significant detection would imply that the stellar tide is not exactly aligned with the planet. The amount by which the stellar tide lags behind the planetary orbit likely depends on the tidal quality factor of the star \citep{pfahl08,barnes10}. $z$ is defined as
\begin{equation}
\cos(z) = -\sin{i} \sin{2\pi\phi}.
\end{equation}
where $i$ is the inclination of the planet with respect to our line of sight with $90^{\circ}$ indicating the planet-star orbital plane is in our line of sight. $i$ is explicitly included in the description of the reflection but is included implicitly in the coefficients $A_{e}$ and $A_{b}$. For the reflection/thermal emission component in the above equation we have assumed a Lambertian phase function \citep{lambert,russell16,sobolev75}. The reason $i$ must be explicitly included in the reflection/emission component is because the shape of the Lambertian phase function depends on the inclination of the planetary orbit whereas inclination effects only the amplitude of the ellipsoidal and Doppler beaming components.

\begin{figure}[th] 
   \centering
      \plotone{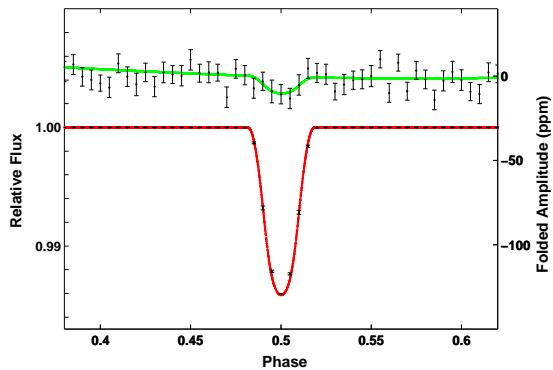}
      \caption{The folded and binned light curve of TrES-2 zoomed in on the transit and the occultation. On top of this we plot the best-fitting photometric model -- red for the transit and green for the occultation. The left-hand scale refers to the transit and the right-hand scale refers to the occultation.}
   \label{fig:transit}
\end{figure}

A best fit model was computed with a Levenberg-Marquardt algorithm \citep{more78,levenberg44,marquardt63}. We show the best-fitting model of the transit and occultation overplotted on the folded data in Figure~\ref{fig:transit}. The best fitting model parameters were then used to seed a Markov Chain Monte Carlo (MCMC)
simulation (an introduction to transit model fitting using MCMC is given in
\citealt{ford06} and \citealt{collier07}) with jumps defined using the Metropolis-Hastings algorithm. We adopted the asteroseismic derived value of $\rho_{\star}$ as a prior. We expect uncertainties on both the observed and modeled light curve data points to be normally distributed and therefore be roughly independent so we describe the likelihood for the model to match the observations as $e^{-\chi^{2}/2}$, where $\chi^{2}$ is the standard chi-squared statistic.
The uncertainty on each data point $\sigma_{f,i}$ is taken directly from the \K{} data files.
We calculated 4 chains of length $10^5$, each with perturbed starting parameters. The first 10\% of chains were disregarded to avoid dependence on the initial starting position. All our parameter chains had a \citet{gelman92} $\sqrt{\hat{R}}$ statistic of $<1$ which indicates good convergence, so we combined the four chains to calculate our final posterior distributions. 

Our model does not include the effect a finite speed of light has on time of occultation relative to the time of transit. This effect is known as R\"{o}mer delay and causes the occultation not to be seen at exactly phase 0.5 for a circular orbit. The R\"{o}mer delay only affects the time of occultation and not the duration of the transit, and hence only the accuracy of $e\cos{\omega}$ is reduced \citep{sterne40,dekort54}. In order to account for this we calculate the time delay we expect \citep{loeb05,kaplan10} and inflate our uncertainty on $e\cos{\omega}$ by converting the time delay into $e\cos{\omega}$ \citep{winn11} and adding in quadrature. We note that the R\"{o}mer delay owing to the motion of bodies in our solar system is explicitly corrected for.

\begin{figure}[th] 
   \centering
      \plotone{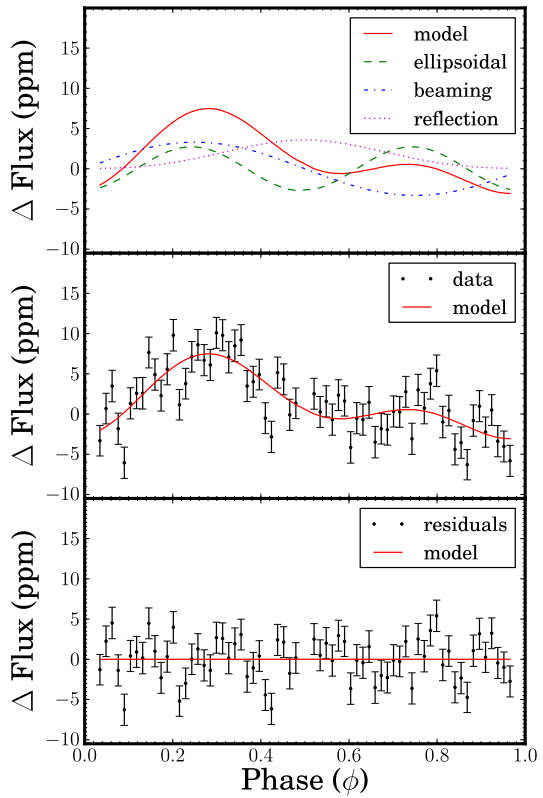}
      \caption{The model phase curve of \Tp{} is shown in the top panel with the three constituent part which make up the model: The ellipsoidal effect (green dashed curve), the Doppler beaming (blue dot-dashed curve) and reflection/emission (purple dotted curve). The sum of these effects is the final model, shown by the red solid line in the top and middle panel. The phase folded \K{} data from Q0--11 is plotted in the central panel. The residuals of this fit are shown in the lower panel. The constant offset in our fits has been removed from the models shown for ease of comparison and the transit and occultation has been cut out. Data has been binned in the figure with 15,000 short cadence data point making up each bin, for ease of interpretation. The fit described in the text was performed on un-folded and un-binned data.}
   \label{fig:phasecurve}
\end{figure}

We do not fit harmonics of the orbital period in order to describe an eccentric orbit \citep[e.g.][]{mislis12a} because in initial transit fitting we found \Tp{} to be on a orbit very close to circular (the upper limit we derive for eccentricity would result in a $<$1\% change in the amplitude, which is below the uncertainty on any amplitude we measure). In Table 2 we report the median of our posterior
distribution. The upper and lower bounds on the uncertainties are the 84.1 and 15.9 percentile of the marginalized posterior distribution, equal to the central 68.2\% of the distribution. The best fitting model of the out-of-transit/occultation effects (using the median of each chain) is shown in Figure~\ref{fig:phasecurve}.

\begin{table*}
\begin{center}
\caption{Planet parameters calculated from the MCMC analysis of the phase curve of \Tp.}
\vspace{0.1cm}
\begin{tabular}{l l l l}        
\hline         
Parameter & Value&Literature value\footnote{Values taken from the literature are provided for comparison. Where the literature values is taken from \citet{kipping11b} we use their eccentric model.}&Reference \\  
\hline
Period (d) &$2.47061320^{+0.00000002}_{-0.00000002}$&$2.47061892^{+0.00000018}_{+0.00000012}$ &\citet{kipping11b}\\
$T_{0}$ (BJD)&$2454955.7625504^{+0.0000051}_{-0.0000060}$&$2454849.526640^{+0.000022}_{-0.000021}$& \citet{kipping11b}\footnote{The transit epoch (T0) given by \citet{kipping11b} differs from ours owing to our fitting being free to choose any transit as the zero-point.}\\
$R_{p} / R_{\star}$&$0.125358^{+0.000019}_{-0.000024}$&$0.1278^{+0.0029}_{-0.0018}$&\citet{kipping11b}\\
b&$0.84293^{+0.00016}_{-0.00015}$&$0.8418^{+0.0037}_{-0.0045}$&\citet{kipping11b}\\
Occultation depth (ppm)&$6.5^{+1.7}_{-1.8}$&$16^{+13}_{-14}$&\citet{kipping11}\\
$e \sin{\omega}$& $-0.00014^{+0.00012}_{-0.00036}$&$-0.009^{+0.0024}_{-0.0029}$&\citet{kipping11b}\\
$e \cos{\omega}$\footnote{The uncertainties on $e \sin{\omega}$ have been inflated to account for R\"{o}mer delay.}& $-0.00016^{+0.00056}_{-0.00067}$ & $0.0005^{+0.0018}_{-0.0018}$&\citet{kipping11b}\\
$A_{e}$ (ppm)& $2.79^{+0.44}_{-0.62}$&$1.50^{+0.92}_{-0.93}$&\citet{kipping11}\\
$A_{b}$ (ppm)& $3.44^{+0.35}_{-0.33}$& $0.22^{+0.88}_{-0.87}$&\citet{kipping11}\\
$A_{r}$ (ppm)& $3.41^{+0.55}_{-0.82}$&$6.5^{+1.9}_{-1.9}$&\citet{kipping11}\\
Flux offset (ppm)&$7.35^{+0.38}_{-0.38}$&--\\
Phase lag ($l$, radians)&$0.01^{+0.37}_{-0.36}$&--\\
Limb darkening coefficients $\{c_{1}, c_{2}, c_{3}, c_{4}\}$&\{0.4330, 0.3552, 0.0450, -0.1022\}& -- \\
\hline
\end{tabular} 
\label{tab:transit} 
\end{center}
\end{table*}

We find significant detections of ellipsoidal, beaming and reflection/thermal emission components in the phase curve. We do not detect any significant phase lag between the stellar tide raised by the planet and the orbital period of the planet. We also fit the three phase curve components independently from the transit and occultation (phases around 0 and 0.5 were cut out), the results were consistent with the joint fit. Additionally, we tried including a $\sin(4\pi\phi)$ component in the phase curve fit as suggested by \citet{faigler11} and measured an amplitude consistent with zero. If the phase variations were due to noise there is no \textit{a priori} reason why we would see positive amplitudes in the three physical components and a zero amplitude for the non-physical component. That the amplitude of a $\sin(4\pi\phi)$ component is consistent with zero supports out conclusion that the phase curve components are real.

In order to further increase our confidence that the phase curve variations are not owing to random noise we compared our model with a model containing no Doppler beaming, ellipsoidal variations or reflection/emission by calculating the Deviance Information Criterion \citep[DIC,][]{spiegelhalter02} for each model. The difference in DIC values indicate that a model containing phase curve variations is favored over the flat model by a factor of $>10^{38}$. A model containing ellipsoidal variations and Doppler beaming but no reflection is disfavored over a model including the reflection by a factor of $4.6\times10^{6}$. Likewise, models containing no Doppler beaming or no ellipsoidal variations but the other two effects are always strongly disfavored over a model with all three effects.

An inclination change of $2\times10^{-4}$ degrees per day has been reported by \citet{mislis09} and \citet{mislis10}, suggesting a secular torque from an otherwise unseen companion.  However, the detection was challenged from the ground by \citet{scuderi10}, and the large magnitude was not confirmed by \citet{schroter12} on the basis of earlier \emph{Kepler} data.  We investigated this question again, now with a longer baseline. 

We fixed $\rho_\star$ to its best-fitting value and allowed $i$ and $R_p/R_\star$ to solve for their best values for each quarter individually.  The motivation for letting $R_p/R_\star$ float is not that we expect it to change physically, but that the quarterly changes in the observations may cause the dilution due to background stars to vary. Thus we guard against misinterpreting a scale change as a duration change due to the rather V-shape lightcurve of this grazing planet. 

The values of $i$ were scattered about the best-fit, with no secular trend.  We measure $di/dt = (1.0 \pm 1.2)\times10^{-6}$~degrees per day, i.e. no change to a 3-$\sigma$ limit of $\sim 4.5\times10^{-6}$~degrees per day.  To interpret this limit in terms of a limit on perturbing planets, we simplify the expression from \citet{ballard10} using the approximation for the Laplace coefficient $b^{(1)}_{3/2} (\alpha) \rightarrow 3 \alpha$ for $\alpha \ll 1$, and find:
\begin{equation}
\begin{split}
M_c &= M_\star \frac{2}{3 \pi} \frac{1}{\Delta \Omega} \frac{P_c^2}{P_b} \frac{di}{dt} \\
  &< 1.2 M_\oplus (\frac{P_c}{10 \rm d})^2 (\frac{\Delta \Omega}{10^\circ})^{-1}.
\end{split}
\end{equation}
where $\Delta \Omega$ is the node of the perturbing planet on the sky plane, relative to the transiting planet, $M_c$ and $P_c$ are the period of the putative perturbing planet, and we have used our limit on $|di/dt|$ in the final inequality.   With this tight limit, we see that we are capable of detecting Earth-mass planets via their secular torques, as envisioned by \cite{miralda02}, but see no evidence for their presence.

\section{Physical parameters from the phase curve of TrES-2b} \label{sec:planet-para}
We use the three amplitudes calculated in our MCMC analysis to determine parameters which describe \Tp.
For radial velocities much lower than the speed of light, the coefficient $A_{b}$ is described by
\begin{equation}
A_{b} = \alpha_{b} \frac{K_{RV}}{c} \: ,
\end{equation}
where $c$ is the speed of light and $\alpha_{b}$ is the photon-weighted bandpass-integrated beaming factor calculated in the manner described by \citet{Bloemen10}, where
\begin{equation}
\alpha_{b} = \frac{\int \epsilon_{K}\lambda F_{\lambda}B\,\mathrm{d}\lambda}{\int \epsilon_{K}\lambda F_{\lambda}\,\mathrm{d}\lambda} \: .
\end{equation}
Here, $\epsilon_{K}$ is the \K{} response function\footnote{The \K{} response function is available in the supplement to the \K{} Instrument Handbook.}, $\lambda$ is the wavelength and $B$ is the monochromatic beaming factor, which \citet{loeb03} give as
\begin{equation}
B = 5 + \frac{\mathrm{d} \ln  F_{\lambda}}{\mathrm{d} \ln \lambda}.
\end{equation}
For \T{}A we compute $\alpha_{b} = 3.97\pm0.03$ using ATLAS model spectra \citep{castelli04} with parameters we derived in \S2.
$K_{RV}$ is the radial velocity semi-amplitude which for a circular orbit is given by
\begin{equation}
K_{RV} = \left(\frac{2\pi G}{P}\right)^{1/3} \frac{M_{p} \sin i}{M_{\star}^{2/3}} \: .
\end{equation}
Here, $G$ is the universal gravitational constant, $P$ is the orbital period of the planet, $M_{\star}$ is the mass of the star, and $M_{p}$ is the planetary mass. Therefore, if we measure $A_b$, we can find the mass ratio of the system.

We adopt the \citet{faigler11} parameterization of $A_{e}$ where
\begin{equation}
A_{e} = \alpha_{e} \frac{M_{p}}{M_{\star}} \left(\frac{R_{\star}}{a}\right)^{3} \sin^{2}{i} \: .
\end{equation}
We calculate $\alpha_{e}$ using a first order approximation of the ellipsoidal variation given by \citet{morris85} and \citet{morris93},
\begin{equation}
 \alpha_{e} =\frac{0.15(15+u)(1+g)}{3-u}
\label{equ:morris}
\end{equation}
where $u$ and $g$ are the linear limb darkening and gravity darkening parameters, respectively. We trilinearly interpolate the limb and gravity darkening coefficients calculated by \citet{claret11} from the grids in effective temperature, surface gravity and metallicity, giving values of $u = 0.580$ and $g = 0.354$.

The geometric albedo is the ratio of the flux observed from a planet, 
compared to that of a perfect Lambert disc with the same radius as that 
of the planet.  While in the solar system this observed flux is 
essentially entirely scattered star light, for hot Jupiters this 
planetary flux can be a mix of scattered light as well as thermal 
emission.  Observationally one cannot determine which is more important 
for a given planet, but with the aid of atmospheric models, the relative 
amounts of scattered and thermally emitted light can be estimated.  We use the amplitude of the reflection/thermal emission component to calculate the geometric albedo, $A_{g}$, for \Tp{} in the Kepler bandpass using the equation
\begin{equation}
A_{r}= A_{g} \left(\frac{R_{p}}{a}\right)^{2}.
\end{equation}

At each step in the Markov chains created in the previous section we calculated the radial velocity semi-amplitude, planet mass, radius and density and geometric albedo. The median values and uncertainties (central 68.27\% of the distribution) on these parameters are given in Table~\ref{tab:phase}. Using the Markov chains enables uncertainties and correlations between parameters to be propagated throughout this work.

\begin{table*}
\begin{center}
\caption{Derived parameters of \Tp. The table is split between those values which are related to the photometric determination of the planet mass but we consider less precise than literature values obtained from radial velocity observations and values obtained from the transit model which exceed that found in the literature.}
\vspace{0.1cm}
\begin{tabular}{l c l l}        
\hline         
Parameter & Value&Literature value&Reference\\  
\hline
Radial velocity semi-amplitude ($m s^{-1}$)&$249^{+24}_{-28}$&$181.3\pm2.6$& \citet{odonovan06}\\
Planet mass from beaming ($M_{\textrm{Jup}}$)&$1.61^{+0.17}_{-0.18}$& --\\
Planet mass from ellipsoidal variations ($M_{\textrm{Jup}}$)&$1.06^{+0.28}_{-0.23}$& --\\
Weighted average planet mass ($M_{\textrm{Jup}}$)& $1.44\pm0.21$& $1.206\pm0.045$& \citet{southworth11}\\
Planet density (g cm$^{-3}$)&$1.14^{+0.12}_{-0.17}$& \footnote{We do not give a density from the literature here owing to the two different sources from which we take mass and radius.}\\
\hline
Planet radius ($R_{\textrm{Jup}}$)&$1.162^{+0.020}_{-0.024}$&$1.187^{+0.034}_{-0.035}$& \citet{kipping11b}\\
$a/R_{\star}$&$7.931^{+0.056}_{-0.046}$&$8.06^{+0.025}_{-0.021}$&\citet{kipping11b}\\
Semimajor-axis (A.U.)&$0.03503^{+0.00053}_{-0.00073}$&$0.03563^{+0.00048}_{-0.000058}$&\citet{kipping11b}\\
Inclination  ($^{\circ}$)&$83.881^{+0.038}_{-0.043}$ &$84.07^{+0.34}_{-0.31}$&\citet{kipping11b}\\
Geometric albedo $A_{g}$&$0.0136^{+0.0022}_{-0.0033}$&$0.0253\pm0.0072$&\citet{kipping11}\\
\hline
\end{tabular} 
\label{tab:phase} 
\end{center}
\end{table*}


\section{Discussion} \label{sec:disc}
\citet{faigler11} derive a parameter $\mathfrak{R}$ to denote the expected ratio of the ellipsoidal amplitude to the beaming amplitude which they define as 
\begin{equation}
\mathfrak{R} \equiv \frac{A_{e}}{A_{b}} = 5 \frac{\alpha_{e}}{\alpha_{b}/4}M_{\star}^{-{4}/{3}} R_{\star}^{3} P_{\textrm{orb}}^{-{5}/{3}} \sin{i}\: ,
\end{equation}
where $M_{\star}$ and $R_{\star}$ are in units of solar mass and solar radius and $P_{\textrm{orb}}$ is in days.
Note that we must divide our value of $\alpha_{b}$ by 4 in order to be consistent with \citet{faigler11}.
For \Tp{} we find $\mathfrak{R} = 0.81^{+0.14}_{-0.13}$, which is  within 2-$\sigma$ of the expected value of $1.34^{+0.19}_{-0.17}$. This difference is not particularly significant and we do not find this overly concerning.

We measure a radius for \Tp{} of $1.162^{+0.020}_{-0.024}$ R$_{\textrm{Jup}}$ which is consistent with previously published values \citep{odonovan06,sozzetti07,daemgen09,southworth11}. However, the radial velocity semi-amplitude we calculate from photometry ($249^{+24}_{-28}$ ms$^{-1}$) is significantly larger than the value measured by \citet{odonovan06} of $181.3\pm2.6$ m s$^{-1}$ using spectroscopic radial velocity data, hence the mass derived from Doppler beaming is also on the high side. Our derived mass is $\sim$1-$\sigma$ higher than the mass found by \citet{odonovan06}, and 2$\sigma$ higher than the mass found by \citet{sozzetti07} who use the same radial velocity data as \citeauthor{odonovan06} but with revised stellar parameters.

Our masses derived from the ellipsoidal variations and from Doppler beaming differ by 2-$\sigma$. The mass derived from beaming is usually taken to be the ground truth because it can be calculated in a relatively model independent way whereas Equation~\ref{equ:morris} is an approximation and relies on the accuracy of the equations used to describe limb and gravity darkening. \citet{shporer11} discuss the limitations of this approximation for a earlier-type star such as KOI-13A, however in the case of a star with a convective envelope such as \T{}A the approximation should describe the amplitude owing to ellipsoidal variations reasonably well. 
For the planet mass we adopt the weighted average of the mass derived from Doppler beaming and from ellipsoidal variations where the weighting is based on the standard deviation the MCMC chain masses as a proxy for uncertainty. This results in the mass of \Tp{} being $1.44\pm0.21 M_{\textrm{Jup}}$. We stress that this value does not supersede previous measurements from spectroscopic radial velocity observations, but merely demonstrates that this technique can be used to determine a planet mass consistent with other observations albeit at lower accuracy. However, the ability to derive masses consistent with valuable multiple epoch, high-resolution spectra for radial velocities should not be overlooked.

As discussed earlier, R\"{o}mer delay is not included in our model. We estimate the observed time of occultation will be offset by 34.5 s compared to what would be seen if the speed of light was infinite \citep{kaplan10,bloemen12}. The  R\"{o}mer delay equates to a change in $e\cos{\omega}$ of 0.00051 \citep{winn11}. We add this value in quadrature to our previously derived uncertainties on $e\cos{\omega}$ to give a final value of $-0.00016^{+0.00056}_{-0.00067}$. This equates to an eccentricity of $0.0002^{+0.0010}_{-0.0002}$. This tight constraint on the eccentricity of the planet (the 3-$\sigma$ upper limit is 0.0025) is consistent with the expectations of tidal dissipation. This low eccentricity suggests that the method of \citet{ragozzine08} for observing apsidal precession due to the planetary interior is not possible with the existing data. 

\begin{figure*}
\plotone{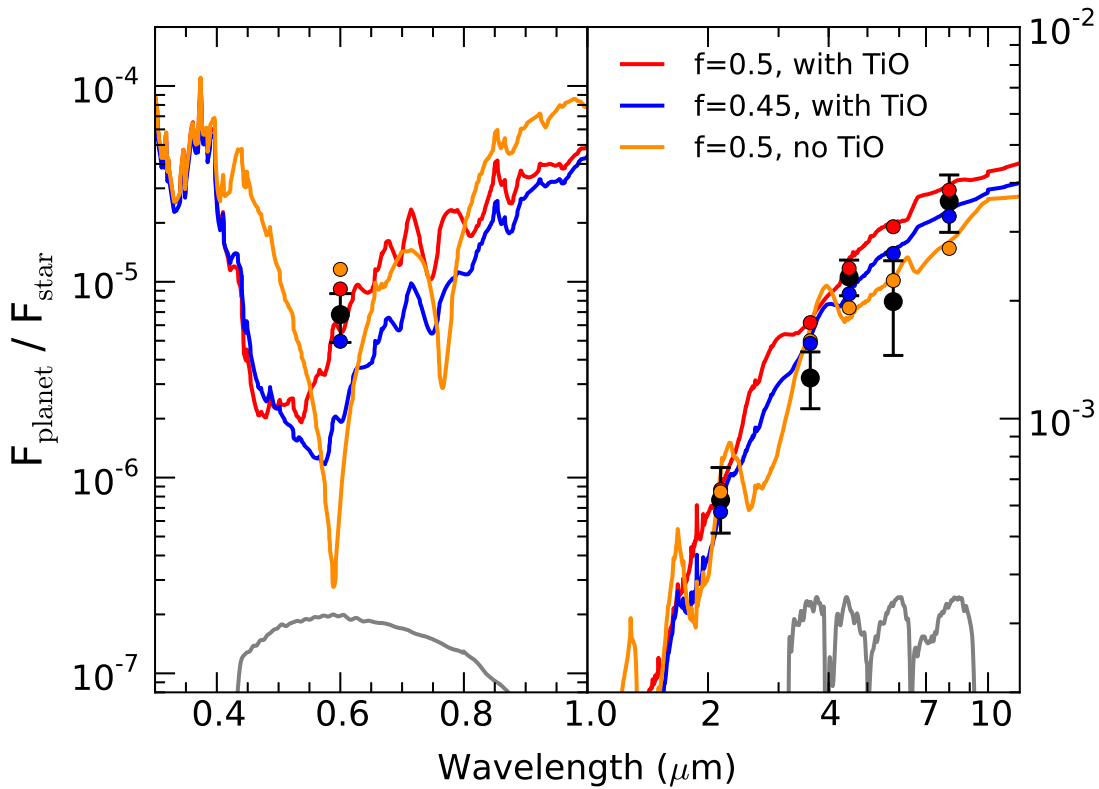}
\caption{\emph{Left panel}: Planet-to-star flux ratios at optical wavelengths.  The transmission function of the \emph{Kepler} bandpass is shown, as well as the measured occultation depth (black, with error bars). \emph{Right panels}:  Planet-to-star flux ratios at infrared wavelengths.  The transmission function of the \emph{Spitzer} IRAC bandpasses are shown, as well as the measured occultation depths (black, with error bars) from \citet{odonovan10} at IRAC wavelengths and \citet{croll10} at $K_{\rm s}$.  The best-fit model is shown in blue, which features a day-side temperature inversion, due to TiO gas, and modest redistribution of absorbed flux to the night side. When fitting models we convolve the bandpass of the observations with the models.
\label{Caroline}}
\end{figure*}

A growing number of planets have infrared emission spectroscopy from Spitzer and adding the Kepler data point to the existing data can be an important extra constraint for differentiating inverted and non-inverted atmospheres \citep{christiansen10}.
We have modeled the atmosphere of TrES-2b using the methods described in \citet{fortney07} and \citet{fortney08}.  In Figure \ref{Caroline} we compare the day-side planet-to-star flux ratio data to three atmosphere models.  The models shown use a redistribution factor $f$ of 0.5 or 0.45, to simulate redistribution of absorbed stellar flux over the entire day side ($f=0.5$) or with a slight loss to the night side $f=0.45$.  The yellow model does not include TiO/VO as gaseous absorbers, leaving gaseous Na and K to absorb optical flux at much higher pressures, and a temperature inversion does not occur.  The blue and red models include equilibrium mixing ratios of TiO/VO, and possess temperature inversions.  TiO and VO gases  have exceptionally strong optical opacity, and can lead to temperature inversions \citep{hubeny03,fortney06}.

The \emph{Spitzer} data plotted in Figure \ref{Caroline} are from \citet{odonovan10} and the $K_{\rm S}$ datum from \citet{croll10}.  Croll et al. used similar methods to Fortney et al. to find a best-fit atmosphere model that included a temperature inversion, but it was only a marginally better fit than the no inversion model.  However, with the inclusion of the \emph{Kepler} data point, a model with a temperature inversion is now more strongly preferred.  \citet{spiegel10} have also modeled TrES-2b, using the same data shown here, except the \emph{Kepler} data point was a preliminary number \citep{kipping11b}.  These authors find a significantly better fit with a temperature inversion as well.  Models that lack TiO or a similar high altitude strong optical absorber are often too bright at optical wavelengths.  This can be seen in Figure \ref{Caroline} and \citet{spiegel10} Figure 2a.  The very low geometric albedo detected for TrES-2b of $0.0136^{+0.0022}_{-0.0033}$ is in line with model predictions for cloud free hot Jupiter atmospheres \citep{sudarsky00}.

The relatively small day-to-night brightness contrast of $3.41^{+0.55}_{-0.82}$ ppm is quite interesting, but potentially difficult to understand in a simple way.  If TiO (or a similar optical absorber) is present on the day side, it may well condense out on the night side \citep[e.g.][]{showman09}.  This could mean the depth probed in the \emph{Kepler} bandpass could be at a pressure $\sim100\times$ larger on the night side.  This same dramatic change in opacity likely does not happen in the near or mid infrared \citep{knutson09}, since H$_2$O is the dominant absorber in both hemispheres, and the day and night abundances of CO and CH$_4$ may be homogenized due to vertical and horizontal mixing \citep{cooper06}.  

A strong possibility for TrES-2b is that the brightness temperature of the inverted day-side atmosphere probes pressures of $\sim$~10 mbar, which may be quite similar in temperature to that on the night side near 1 bar.  Another possibility for the warm night side of TrES-2b could be energy dissipation due to a variety of potential mechanisms \citep{showman02,arras10,batygin10,youdin10}, which could be expressed as a higher temperature at 1 bar. We note that \citet{welsh10} also found a small ($\sim$~300 K) day-night brightness temperature contrast for HAT-P-7b in the \emph{Kepler} band, although \citet{jackson12}, with more data, suggest the difference is greater than 700 K. 

The day-side to night-side planet flux ratio we determine is lower than that found by \citet{kipping11}who detected a phase curve reflection/emission component with a full-amplitude of $6.5\pm1.9$ ppm which is higher than the $3.41^{+0.55}_{-0.82}$ ppm we detect, although still within 1--2-$\sigma$.

\section{Summary}
We have discovered ellipsoidal variations and Doppler beaming in the phase curve of the \T{} star-planet system. An asteroseismic analysis of the solar-like oscillations on the star allows us to derive an accurate stellar mass and radius. Using the approach of \citet{faigler11} we derived a mass of \Tp{} of $1.61^{+0.16}_{-0.18}$ $M_{\textrm{Jup}}$ from the Doppler beaming effect and $1.06^{+0.28}_{-0.23}$ $M_{\textrm{Jup}}$ from the ellipsoidal variations. That these two values are independent and that they agree to within 2-$\sigma$ and are within in 1--3$\sigma$ of literature values strengthens our belief that the variations are not instrumental or induced by stellar variability. We are further convinced that what we detect is truly induced by the planet because the ratio of ellipsoidal to beaming amplitudes agrees within $\sim$2-$\sigma$ of theoretically expected values.

We detect a difference between the day and the night side flux from \Tp{} of the level of $3.41^{+0.55}_{-0.82}$ ppm, lower than the amplitude reported by \citet{kipping11}. This suggests that \Tp{}, the darkest known exoplanet, is even darker than previously thought.

\acknowledgments
This paper includes data collected by the \K{} mission. Funding for the \K{} mission is provided by the NASA Science Mission directorate. All \K{} data presented in this paper were obtained from the Mikulski Archive for Space Telescopes (MAST) at the Space Telescope Science Institute (STScI). STScI is operated by the Association of Universities for Research in Astronomy, Inc., under NASA contract NAS5-26555. Support for MAST for non-HST data is provided by the NASA Office of Space Science via grant NNX09AF08G and by other grants and contracts.
SB acknowledges funding from the
European Research Council under the European Community's Seventh Framework Programme
(FP7/2007--2013)/ERC grant agreement n$^\circ$227224 (PROSPERITY), as
well as from the Research Council of KU Leuven grant agreement GOA/2008/04. We thank Simchon Faigler and the Kepler Science Team for providing insightful comments during the preparation of this manuscript.
DH is supported by appointment to the NASA Postdoctoral Program at Ames Research Center, administered by Oak Ridge Assocaited Universities through a contract with NASA.

\bibliographystyle{apj}

\end{document}